\documentclass[sigconf]{acmart}

\def\BibTeX{{\rm B\kern-.05em{\sc i\kern-.025em b}\kern-.08emT\kern-.1667em\lower.7ex\hbox{E}\kern-.125emX}}
    
\copyrightyear{2019}
\acmYear{2019}
\setcopyright{acmlicensed}
\acmConference[]{}{}{}
\acmBooktitle{}
\acmPrice{}
\acmDOI{}
\acmISBN{978-1-4503-9999-9/18/06}
\usepackage[caption = false]{subfig}
\usepackage{mathtools}
\begin{document}

%
\title{Modeling and Counteracting Exposure Bias in Recommender Systems}

%
\author{Sami Khenissi}

\email{sami.khenissi@louisville.edu}
\author{Olfa Nasraoui}
\email{olfa.nasraoui@louisville.edu}
\affiliation{%
  \institution{University of Louisville}
}

%

%
\begin{abstract}
What we discover and see online, and consequently our opinions and decisions, are becoming increasingly affected by automated machine learned predictions. Similarly, the predictive accuracy of learning machines heavily depends on the feedback data that we provide them. This mutual influence can lead to closed-loop interactions that may cause unknown biases which can be exacerbated after several iterations of machine learning predictions and user feedback. Machine-caused biases risk leading  to undesirable social effects ranging from polarization to unfairness and filter bubbles.

In this paper, we study the bias inherent in widely used recommendation strategies such as matrix factorization. Then we model the exposure that is borne from the interaction between the user and the recommender system and propose new debiasing strategies for these systems.

Finally, we try to mitigate the recommendation system bias by engineering solutions for several  state of the art recommender system models.
 
 Our results show that recommender systems are biased and depend on the prior exposure of the user. We also show that the studied bias iteratively decreases diversity in the output recommendations.  Our debiasing method demonstrates the need for  alternative recommendation strategies that take into account the exposure process in order to reduce bias.
 
 Our research findings show the importance of understanding the nature of and dealing with bias in machine learning models such as recommender systems  that interact directly with humans,  and are thus causing an increasing influence on human discovery and decision making. 
 
\end{abstract}

%
%

%

%

%
\maketitle

\section{Introduction}
Machine learning models make several assumptions in order to provide unbiased predictions. One of these assumptions is the fact that the data is collected randomly. Usually, in a machine learning project, human experts are responsible for collecting and labeling the data. For example, a classic problem of building a machine learning model that can classify images of dogs and cats starts with collecting labeled data of images. Images of cats and dogs are collected randomly and labeled by an expert or an oracle to create labeled training and testing data. 

Recommender systems should be treated differently because of the way the data is collected. Indeed, human behavior is responsible for most of the data collection and labeling. This happens because most recommender systems collect usage or interest data such as views, clicks or ratings from online users in order to make future predictions. The users then see the recommendations and, through their interaction, provide the next batch of ratings. This closed feedback loop generally narrows the available data to only the items that the user has been exposed to. To have an unbiased system, the data shown to the user should be randomly selected and then the user should rate all the seen items (to eliminate the user bias). However this is far from being realistic. This process causes several serious problems for both the users and the items. For instance, the user may experience a filter bubble problem. When the model fails to learn all of the users' diverse interests, it can keep providing the same types of recommendations again and again. This can also cause polarization \cite{Chaney:2018:ACR:3240323.3240370}. Furthermore, items in a recommender system may suffer from underexposure. The closed feedback loop can cause an unfair exposure between the different items (movies, books, articles...) which may result in a skewed rating distribution and thus a minority of popular items and a large number of unpopular or underexposed items. 

In this paper, we ask the following questions: 1) What is the asymptotic behavior of the generic recommender system in terms of recommending new items  2) How can we model the user exposure in an iterative closed feedback loop? 3) Can we reduce the exposure bias effect in a collaborative filtering setting? 

This paper presents the following contributions:
\begin{itemize}
    \item Presenting an experimental framework to test exposure models
    \item Developing a new debiasing strategy that can mitigate the effect of exposure bias on collaborative filtering recommendations models
    \item Developing an experimental protocol to simulate the feedback loop between users and the recommender model, track its effect using an existing real-life dataset and evaluate debiasing strategies.
\end{itemize}

\section{Related Work}
Several models have emerged trying to counteract the exposure bias problem. Dealing with the different kinds of bias can be done in several stages of the recommender system's pipeline. Some methods use post learning techniques which consist of different ranking and selection strategies after performing the predictions. In this family, we find a big interest in Multi-Armed Bandit techniques \cite{2019arXiv190210730J} since they have proven to be efficient in the exploitation vs exploration problems \cite{Katehakis1987}. 
MAB are good with exploration problems, they are often used in Recommender systems in order to solve the exposure bias problem. For instance, a group of DeepMind researchers \cite{2019arXiv190210730J} have recently used MAB to study the effect of the feedback loop in recommender systems. They showed that the feedback loop can decrease the quality of the recommendations and they also showed that random exploration using Multi-Armed Bandit techniques can enhance and boost the quality of the predictions.

Other approaches  \cite{DBLP:journals/corr/abs-1901-07555} used other ranking strategies such as accounting for the diversity while providing recommendations. 
Other techniques focused on eliminating the mathematical bias of the learning algorithm \cite{DBLP:journals/corr/SchnabelSSCJ16}. Schnabel et al \cite{DBLP:journals/corr/SchnabelSSCJ16} provided an inverse propensity strategy that eliminates the bias of the mathematical estimator for the training error and provides an unbiased estimation of the training performance.  The advantage of this method is that it takes into account the confounding factor resulting from the exposure bias.  
Additional work used regularization to account for the bias in the data. Abdollahpouri et al \cite{Abdollahpouri:2017:CPB:3109859.3109912} provided a regularization strategy that accounts for the popularity in a recommender system. The main idea of the model is to push recommendation in a way that balances the accuracy along with the intra-list diversity. 

Some work focused on the iterated bias of Recommender Systems. Bountouridis et al. \cite{Bountouridis:2019:SSF:3287560.3287583} designed a simulation framework to see the effect of the recommendation models on the diversity and novelty of the Recommendations. They used news data and they compared different state-of-the-art models. 

\cite{2019arXiv190210730J} also provided a simulation framework \cite{2019arXiv190210730J} where they showed that the iterative effect of the recommender systems decreases the utility of the recommendations. 

Sun et. al \cite{weninproceedings} presented simulations to study the effect of the feedback loop from a machine learning perspective. They used synthetic data and hypothesis testing in order to study how the predictions shift as a result of interactions between the human and the model. \cite{sami} presented a study of several exposure bias counteraction strategies within an iterated framework.

\section{Methodology}
\subsection{Notation}
In the next sections we define $U$ as the set of all users and $I$ as the set of all items. $E$ is the exposure matrix, $P$ and $Q$ are the representation of users and items in the latent space, $\alpha$ is the learning rate, $\beta$ is the regularization rate, $K$ is the dimension of the latent space, $\lambda$ is the Popularity and Exposure Aware Regularization hyperparameter, $JSD$ is the Jensen Shanon Divergence and $R_{ui}$ is the rating given by user $u$ to item $i$. If the rating is missing we set $R_{ui} = 0$.

\subsection{User Exposure distribution}
We define the user exposure distribution as the probability that the user $u$ has seen item $i$. This distribution defines the  likelihood that a given user has been exposed to different items. We define this distribution as a matrix $E$ mapping the users to the items where:
\begin{equation}
    E_{ui} = P(\text{user }u \text{ has seen item } i)
\end{equation}

This distribution is also called propensity \cite{DBLP:journals/corr/SchnabelSSCJ16}, and hard to estimate because it depends on random factors such as the demographic properties of the users. Users from different ages, locations, etc, are exposed to a different set of items. Propensity also depends on the popularity of the item. Popular items are more likely to be seen than other items. It also depends on the websites, social networks, and different platforms that the user may have used since they can promote different ads and products. Another important factor that affects user exposure distribution is the previous iteration of recommendations. In fact, the recommender system is exposing the user to a new set of items with each recommendation. Items with a high likelihood of being recommended will have higher chance of being seen.

The exposure distribution is thus affecting the data collection process. In fact, the user cannot rate an item if he/she has not seen it, hence the need to study this distribution and understand how we can use it in order to mitigate the resulting exposure bias.

\subsubsection{Fair Exposure}
We define the fair exposure distribution as the uniform distribution. In fact, a fair exposure across all items means that the user has equal chances of seeing all the items.
\begin{equation}
    P(\text{user }u \text{ has seen item } i) = \frac{1}{|I|}
\end{equation}

An example of a fair exposure is a recommender system that is starting with totally new items (the users have never seen any of the items in the recommender system). Then the recommender system will recommend a set of items randomly (based on the uniform distribution). This way, the user will have equal chances of seeing and rating these items. The resulting trained model will be unbiased since we used an unbiased data collection strategy in order to get the ratings. However, this scenario is unfeasible when dealing with a real-world setting. The main purpose of a recommender system is to provide relevant recommendations. We aim to develop a method that keeps providing accurate predictions and at the same time takes into account the different exposures for each user and item.

\subsubsection{Popularity Based Exposure}
Some previous work used item popularity \cite{Castells2015} \cite{Steck:2011:IPR:2043932.2043957} as an estimator for the exposure distribution. 
The popularity based exposure model is defined by the following equation:

\begin{equation}
    E_{ui} = \frac{|\{ R_{ui} \neq 0 \}|}{|U|}
\end{equation}

It is important to note that the popularity based exposure model depends only on the item and not the user. It cannot provide a personalized estimation of the exposure for each user. 

\subsubsection{Poisson Factorization based model}

Popularity based exposure models cannot provide personalized approximations for each user. \cite{DBLP:journals/corr/SchnabelSSCJ16} suggested using learned models such as Naive Bayes or logistic regression. In this work, we are going to test another probabilistic framework that is more suitable for the recommender system setting, namely Poisson Matrix Factorization \cite{Gopalan:2014:CRP:2969033.2969181},\cite{63424a9543a64bd3b0296abf8d5c44d9}, \cite{DBLP:journals/corr/GopalanHB13},\cite{Liang:2016:MUE:2872427.2883090}. 

\subsection{Popularity and Exposure Aware Regularization for Matrix Factorization (PEAR-MF)}
 
 As shown in section 3, a recommender system that minimizes a given distance between seen items and unseen items is prone to producing filter bubbles. Matrix Factorization for instance is using the cosine distance through the dot scalar of the latent space representations.
 
 To weaken the assumption on the matrix factorization, we propose a new regularization function based on the exposure model in order to make matrix factorization aware of the exposure bias.

The new objective function for PEAR-MF is:
\begin{equation}
\begin{multlined}
    J(P,Q) = \sum_{u \in U ,i \in I} (R_{ui} - P_uQ_i^T)^2 + \beta (||P_u||^2 + ||Q_i||^2) + \\
    \lambda JSD_u(E_u||\frac{1}{|I|}) (||P_u||^2 + ||Q_i||^2)
\end{multlined}
\end{equation}

$JSD$ is the Jensen Shannon Divergence \cite{inproceedingsjsd}. It is a statistical distance that measures the similarity between two distributions. 

\begin{equation}
    JSD(D_1,D_2) = \frac{1}{2}D_{KL}(D_1||M) + \frac{1}{2}D_{KL}(D_2||M)
\end{equation}

Where
\begin{equation}
    M= \frac{1}{2}D_1 + \frac{1}{2}D_2
\end{equation}
and $D_{KL}$ is the Kullback-Leibler divergence given by
\begin{equation}
    D_{KL}(D_1||D_2) = -\sum_{x \in \chi}E(x)\log(\frac{D_1(x)}{D_2(x)})
\end{equation}

In the new regularization function, we calculate the JSD between the exposure distribution E and the uniform distribution for each user. If the JSD is equal to zero, then this means that the user had a fair exposure to all the items, which means that the use of the regular matrix factorization will not affect the predictions. 

If the JSD is close to one, then the estimated exposure distribution is maximally dissimilar from the uniform distribution. This means that the user has experienced under-exposure or over-exposure to certain items. In this case, the regularization term will contribute to penalizing the error function so that the algorithm adjusts its parameters to fit the new information. Without the exposure related term in equation (12), the regularization treats all users and items equally when trying to reduce overfitting. With the exposure term, the amount of regularization is modulated in proportion to the extremeness of the exposure bias. This shows that the exposure bias problem and the prediction accuracy are related.

The algorithm uses the Alternating Least square optimization method. It alternatively updates P and Q until we reach a fixed number of iterations or convergence. The update equations based on Gradient Descent in each iteration $(t)$ are as follows 

\begin{equation}
    P_u^{t+1} = P_u^t - 2\alpha (R_{ui}-P_u^tQ_i^T)Q_i - 2\beta P_u^t - 2\lambda JSD_u(E_u||\frac{1}{|I|}) P_u^t
\end{equation}
\begin{equation}
    Q_i^{t+1} = Q_i^t - 2\alpha (R_{ui}-P_uQ_i^{tT})P_u - 2\beta Q_i^t - 2\lambda JSD_u(E_u||\frac{1}{|I|}) Q_i^t
\end{equation}
The proposed regularization function can be applied to other algorithms that are used to predict the ratings of the user and trained using gradient descent. Its main idea is to include the information of the over-exposure or under-exposure problem of a given user by increasing the error of the optimization process.

\section{Experimental Results}

For the experimental results we used the Movie-lens dataset \cite{Harper:2015:MDH:2866565.2827872} with 100K ratings and 1M ratings. The ratings in both datasets range from 1 to 5 and 0 is given for a missing rating. All the experiments were repeated 10 times.

In our work, we focus on simulating the feedback loop in recommender systems. This is a very crucial part because we do not want to only evaluate the offline performance of our algorithm, but also to evaluate its behavior in an iterative framework which is closer to the real world performance.

However, this approach comes with limitations in terms of the required dataset. In order to get an accurate performance of the model in an iterated way, we need to know the complete ratings of the user including the missing ratings. For this reason, we adopt a simple trick consisting of generating a semi-synthetic dataset as was done in Schnabel et al. \cite{DBLP:journals/corr/SchnabelSSCJ16}. 
After creating the complete matrix, we proceed to simulating the feedback loop of our recommender system. We train our model using the initial set of training data, then we select 10 recommendations to present to the user. At this step, we use these recommendations for calculating the performance metrics. Then we select a subset from these recommendations in order to add it to the training set based on the relevance of each item from the complete matrix.

\subsection{Modeling the exposure bias}

\begin{figure}
\centering    
{\includegraphics[width=\linewidth]{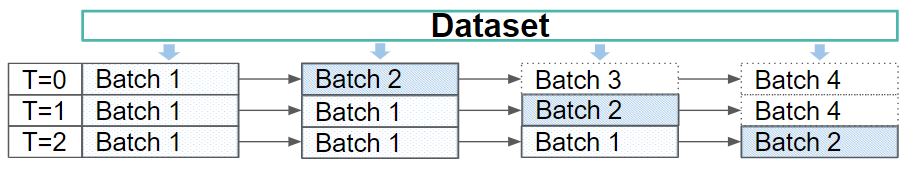}}
\caption[Experimental protocol for evaluating exposure bias]{Example of the learning and Testing process: dotted boxes are the portion of the data used for training and the hatched box is the portion used for testing.}
\label{fig:split}
\end{figure}

The next part of our experiments will verify how good an approximation our exposure model is. We aim at classifying each item as seen or not seen using the known interaction (user, item) in the available data. We use a sliding window as shown in Figure \ref{fig:split} where we split the data into four batches respecting the timestamp Then we iteratively train our models on one part and test on the next one as shown in Figure \ref{fig:split} and then we average the performance. To evaluate the performance of the models Area Under curve (AUC) is used.

\begin{table}
  \caption{Comparison of Poisson Factorization and Popularity Model to predict the exposure of the user on the 1M dataset}
  \label{tab:freq}
  \resizebox{0.25\textwidth}{!}{%
  \begin{tabular}{ccl}
    \toprule
    Model& AUC\\
    \midrule
     \textbf{Poisson Factorization} & \textbf{0.861 $\pm 10^{-4}$  }\\
     Popularity model &  0.815 $\pm 10^{-4}$ 
\\
  \bottomrule
\end{tabular}
}
\label{tab:ev}
\end{table}
Table \ref{tab:ev} shows that Poisson factorization based exposure significantly outperforms the popularity based model with a p-value $< 10^{-3}$. This is in line with our expectations because Poisson factorization learns a more personalized model for user exposure compared to the popularity-based model. The popularity model differs only from item to item but it does not differ from user to user.
For the rest of our experiments, we will consider Poisson factorization as our exposure model

\subsection{PEAR-MF: Counteracting exposure bias}

To evaluate the performance of our algorithm we use the following metrics:
\begin{itemize}
    \item Expected Novelty (EPC) \cite{Castells2015} is a measure that evaluates the expected number of relevant items not previously seen by the user. $EPC = C \sum_{i_k \in Rc} disc(k) p(rel | i_k,u) (1- p(seen|i_k))$
    \item Expected Diversity (EPD): \cite{Castells2015} This metric measures the amount of diversity in each recommendation list based on the pairwise distance between the items in the recommendation and the items that the user has already interacted with. $   EPD = C' \sum_{i_k \in Rc, j \in I_u} disc(k) p(rel | i_k,u)p(rel | j,u) d(i_k,j)$
    \item Gini Coefficient\cite{gini1912variabilità} is used to calculate the balance within the rating distribution. 
    \item Hit Rate defines the percentage of items that the user will interact with from all the items in the recommended list. It captures the quality of the recommendations
\end{itemize}
To test the performance of our model we compare our model to Matrix Factorization \cite{Koren:2009:MFT:1608565.1608614}, Propensity-MF \cite{DBLP:journals/corr/SchnabelSSCJ16} and also mixed strategies using MAB (Naive Multi-Armed Bandits Strategy + MF and Naive Multi-Armed Bandits Strategy + PEAR-MF). Parameters are tuned using 5-fold cross-validation and the best values are selected: $\alpha = 0.001$, $\beta = 0.01$, $\lambda =1$, $K = 10$.
Furthermore, experiments are repeated 10 times and the $95\%$ CI is calculated and a t-test is performed to assess the statistical significance of the results.
\begin{figure}
\centering    
\subfloat[EPC evolution for MovieLens 100K]{\includegraphics[width=0.5\linewidth]{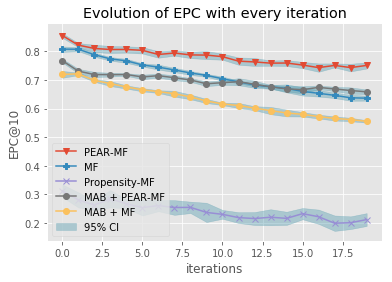}}
\subfloat[EPC evolution for MovieLens 1M]{\includegraphics[width=0.5\linewidth]{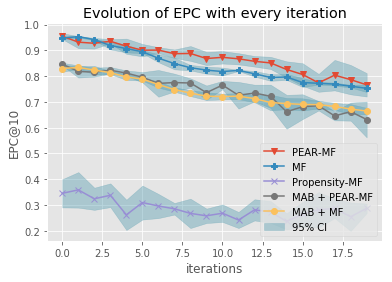}}\\

\subfloat[EPD evolution for MovieLens 100K]{\includegraphics[width=0.5\linewidth]{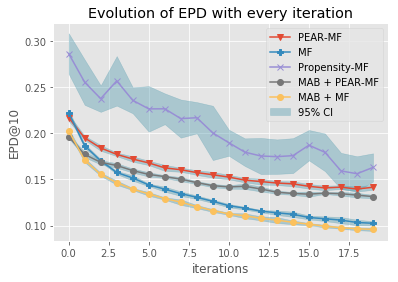}}
\subfloat[EPD evolution for MovieLens 1M]{\includegraphics[width=0.5\linewidth]{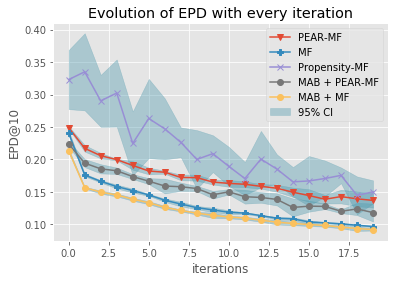}}\\

\subfloat[Gini evolution for MovieLens 100K]{\includegraphics[width=0.5\linewidth]{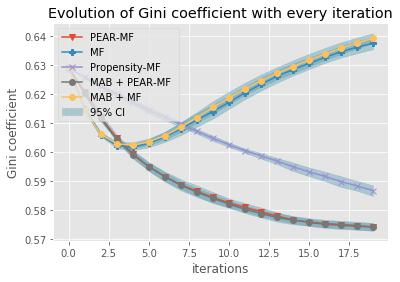}}
\subfloat[Gini evolution for MovieLens 1M]{\includegraphics[width=0.5\linewidth]{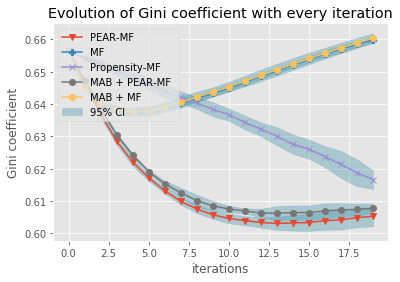}}\\

\subfloat[HR evolution for MovieLens 100K]{\includegraphics[width=0.5\linewidth]{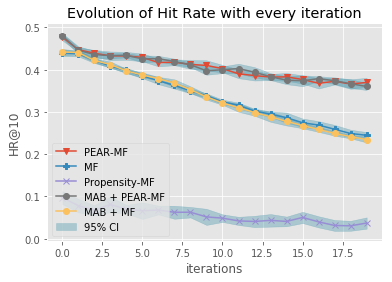}}
\subfloat[HR evolution for MovieLens 1M]{\includegraphics[width=0.5\linewidth]{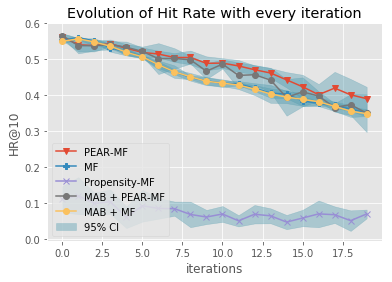}}

\caption{Evaluation Metrics and their 95\% CI tracked in an iterative way. PEAR-MF provides more novel results (a,b) and promotes items that are underexposed (e,f). Furthermore, the Hit Rate performance (g,h) shows that PEAR-MF is providing better quality recommendations. Propensity-MF is providing less accurate but more diverse recommendations (c,d)}
\label{fig:evaluation}
\end{figure}

Figure \ref{fig:evaluation} (a,b) shows that PEAR-MF significantly (p-value $< 10^{-6}$) outperforms the other strategies in terms of Expected Novelty. It proves that the recommendations provided by PEAR-MF are relevant and also have a low probability of being seen by the user. It is also important to note that EPC is decreasing with every iteration. This confirms the feedback loop effect of recommender systems and how it contributes to creating filter bubbles. We also see that MAB strategies do not help at improving the results because they decrease the relevance of the recommendations. Propensity-MF has a lower performance than other methods which is probably due to a low relevance of the recommendations as confirmed by Figure \ref{fig:evaluation} (g,h).

Figure \ref{fig:evaluation} (c,d) shows that Propensity-MF is significantly better than MF and PEAR-MF at providing diverse recommendations. This is due to the fact that Propensity-MF favors items with low exposure, hence the items will be more diverse. However, the results of PEAR-MF seem to have a large variance and decrease with every iteration. Propensity-MF still outperformes (p-value $< 10^{-5}$) MF by providing more diversified predictions.

Figure \ref{fig:evaluation} (e,f) shows how PEAR-MF improves the balance in the rating distribution. After each iteration, the Gini coefficient is decreasing which proves that this method is recommending more items from the long tail. It also outperforms the regular Matrix Factorization model (p-value $< 10^{-7}$).
Propensity-MF also contributes to balancing the rating distribution by decreasing the Gini coefficient after each iteration.

Figure \ref{fig:evaluation} (g,h) shows that PEAR-MF and MAB + PEAR-MF are providing more relevant recommendations than MF (p-value $< 10^{-6}$). This proves that fixing the exposure bias helps improving the quality of the recommendations. In fact, in an iterated framework, the accuracy may be considered good  (inline with the user's taste) but the quality can be low. For instance, a user who keeps seeing the same type of recommendation through several iterations can lose interest and stop interacting with these recommendations.

\section{Conclusion}
In this paper, we provided a personalized model to model the user exposure along with the experimental protocol used to confirm its effectiveness. We designed a new regularization model that can mitigate the exposure bias problem. Finally, we showed the effect of the feedback loop by running simulations that mimic the life cycle of a Recommender System and showing how to evaluate the extent of the exposure bias problem.

Our work has a few limitations. We used semi-synthetic data in order to be able to run simulations. A user study would be more accurate. Also, we worked only with movie data. Other types of data should be investigated such as news and social media interactions.

Because recommender systems increasingly control what humans discover, unbiased recommendations promise to improve fairness and expand the human discovery potential.

\section*{ACKNOWLEDGEMENTS}

This work was supported by National Science Foundation grant NSF-1549981. 


%
\bibliographystyle{ACM-Reference-Format}
\bibliography{sample-base}

%

\end{document}